\documentclass[aps,twocolumn,eqsecnum]{revtex4}
\usepackage{graphics}

\begin{document}

\title{Phenomenological constraints on the Jaffe-Wilczek model
of pentaquarks}

\author{Thomas D. Cohen} \email{cohen@physics.umd.edu}

\affiliation{Department of Physics, University of Maryland,
College Park, MD 20742-4111}

\begin{abstract}
A model recently introduced by Jaffe and Wilczek based on the
quarks being dynamically bound into diquarks has been used to
predict that the recently observed exotic baryons (pentaquarks)
fall into a nearly ideally mixed combination of the 8 and
$\overline{10}$ representations of SU(3) flavor.  The model
predicts two states with nucleon quantum numbers which have
tentatively been identified with the ${\rm N}^*(1440)$ and the
${\rm N}^*(1710)$. This paper examines the viability of this
model by focusing on the decay width of the nucleon members of
the multiplet.  An inequality relating the partial widths of
these nucleon states in the $\pi$+nucleon channel to the width of
the $\theta^+$ is derived for this model under the assuming ideal
mixing  and that the only significant exact SU(3) symmetry
violations are the result of ideal mixing, threshold effects and
the masses of pseudo-Goldstone bosons. This inequality is badly
violated if the states in the multiplet are the ${\rm N}^*(1440)$
and the ${\rm N}^*(1710)$ and if the recent bounds extracted for
the $\theta^+$ width are reliable. Thus, the model appears to
require a scenario with the existence of at least one presently
unknown resonance with nucleon quantum numbers.
\end{abstract}


\maketitle

\section{introduction\label{Intro}}
There is considerable ferment in the field of hadronic physics. A
narrow baryon resonance with a strangeness of +1 (the $\theta^+$)
has been identified by a number of experimental groups\cite{exp}.
This state is clearly exotic in the sense that it cannot be
described by a simple quark model.  There is also experimental
evidence for an additional exotic state with $S=-2$ and $Q=-2$
\cite{exp2}. Almost as remarkable as the exotic nature of these
states is their extremely small widths. All measurements of the
widths are consistent with experimental resolution of $\sim 20$
MeV\cite{exp,exp2} and comparisons with previous experiments
bound the width of the $\theta^+$ to be very
small\cite{width1,width2,width3,width4,width5}; recent analyses
of these experiments suggest that the  width  is less than 1-2
MeV\cite{width2,width3}. The question of how to interpret such
states has become pressing.

Of course, in one sense the interpretation is simple---such states
must emerge directly from QCD.  Early attempts at doing this via
lattice methods have been made\cite{Lat}.  Whether the state of
lattice technology is presently good enough so that the various
extrapolations in quark mass, lattice spacing and lattice volume
are under control for observables associated with these exotic
states are reliable remains to be seen.  In any event, it is
plausible that the lattice will ultimately be capable of
determining the $\theta^+$ mass reliably.  However, even if one
ultimately has a reliable lattice determination of the $\theta^+$
mass it cannot truly be said that one understands the nature of
the exotic states; one merely knows that the state arises from
QCD. Understanding requires some intuitive picture which captures
the essence of the physics, {\it i.e.} a model, along with some
understanding of how the model is connected with the underlying
theory.  In this sense, the simple interpretation that the state
emerges from QCD, though correct, is incomplete. An analogy to
condensed matter physics is useful: the BCS theory of
superconductivity is intrinsically limited but it allows one to
abstract the essential physics far more readily than does a full
microscopic many-body theory.

Thus, there have been multiple attempts to find simple models
which capture the core of the physics associated with exotic
baryons.  One class of models  are the chiral
solitons\cite{ChiSol,IKOR} which have the virtue of being
connected with large $N_c$ QCD and to approximate chiral symmetry
in a manner similar to that of QCD.  Unfortunately most of the
analyses using such models has been based on rigid-rotor
quantization\cite{ChiSol} which is highly
controversial\cite{coh1,IKOR,Pob,DP,coh2}.  It is the view of this
author that the arguments in refs.~\cite{coh1,IKOR,Pob,coh2} are
compelling and that rigid rotor quantization is almost certainly
wrong in that it has no justification from large $N_c$ QCD.
Objections to this view expressed in ref.~\cite{DP} were fully
rebutted in ref.~\cite{coh2}.  Thus, all of these treatments of
exotic baryons in chiral soliton models cannot be considered
valid. There has been an attempt\cite{IKOR} to describe the
exotic states in a consistent manner in these models using the
Callen-Klebanov method\cite{CK}. However, such an attempt is
highly dependent on details of the interaction which are hard to
constrain from other phenomenology; therefore such an approach
remains largely unpredictive.

Another class of models are variants of constituent quark models
\cite{JW,qm,SZ}.  Perhaps most prominent among these is a model
introduced by Jaffe and Wilczek\cite{JW} which describes the
exotic states as pentaquarks composed of two strongly bound
scalar diquarks interacting with an anti-quark.  The model
predicts that the lowest-lying pentaquark states are a nearly
ideally mixed combination of an SU(3) flavor octet and
anti-decuplet.  One key feature of this model is the prediction
of two pentaquark states with nucleon quantum numbers: one with a
mass of $\sim 1450$ MeV and the other with a mass of $\sim 1700$
MeV.  These states have been tentatively identified with the ${\rm
N}^*(1440)$ (also referred to as the Roper resonance) and the
${\rm N}^*(1710)$. The purpose of the present paper is to explore
the phenomenology of the Jaffe-Wilczek model.   In particular, it
focuses on the widths of the $N^*$ states with emphasis on
determining whether the  model is consistent with presently known
phenomenological constraints. It will be shown that the
identification of the two nucleon states as the Roper resonance
and the ${\rm N}^*(1710)$ is inconsistent with the phenomenology
of the widths given the assumptions underlying the model and
recent bounds on the  $\theta^+$ width.

The reason for this inconsistency in the identification is easy to
see. Given the assumption of nearly ideal mixing, the two physical
states with nucleon quantum numbers each have substantial
contributions from both the octet and anti-decuplet  multiplets.
The  $\theta^+$, which has no octet contribution, is extremely
narrow. Hence the coupling of the anti-decuplet part of the two
states to the pion plus nucleon final state is quite small. In
contrast, the Roper resonance is known to be very broad with a
large coupling to the pion plus nucleon state.  If the Roper
resonance is identified with the lighter of the two ${\rm N}^*$
states in the ideal mixed octet plus anti-decuplet, one can
deduce that the coupling to the octet component must be large.  A
large coupling to the octet, in turn, implies a large partial
width for the higher of the two ${\rm N}^*$ states.  However, the
partial width for the ${\rm N}^*(1710)$ into pion plus nucleon
is, in fact, quite small.

In this paper, the simple argument presented above will be
quantified. In particular, an inequality will be derived relating
the partial widths for the decay of the two ${\rm N}^*$ states
into $\pi$ + $N$ with the width of the $\theta^+$. The only
assumptions entering the inequality are that the physical states
are ideally mixed, and that the only sources of SU(3) violation
are ideal mixing, threshold effects and masses of
pseudo-Goldstone bosons. These assumptions are not quite correct:
the Jaffe-Wilczek model is based on {\it nearly} ideal mixing and
other sources of SU(3) violation are present. Thus the inequality
may be modestly violated even if the model is essentially
correct; however, gross violations of the inequality are simply
not compatible with the model. Identifying the pentaquark ${\rm
N}^*$ states as the Roper resonance and the ${\rm N}^*(1710)$ and
putting in the various widths, one finds that the inequality is
grossly violated provided the $\Theta^+$ width is as narrow as
reported in  refs. \cite{width2,width3}.

If one accepts  the validity of the model and the reported bounds
on the $\theta^+$ width then one of three scenarios must be
adopted: i) the ${\rm N}^*(1710)$ state is, in fact, a nearly
ideally mixed pentaquark and it has a previously undiscovered
narrow partner with a mass in the neighborhood of $\rm 1450$ MeV;
ii) the Roper resonance is a nearly ideally mixed pentaquark and
it has a previously undiscovered broad partner with a mass in the
neighborhood of $\rm 1700$ MeV;  iii) neither the lower the lower
nor the upper ${\rm N}^*$ states have been detected.

This paper is organized as follows:  In the following section the
Jaffe-Wilczek model is discussed.   Section (\ref{inequality})
derives an inequality for the partial widths of the ${\rm N}^*$
states based on the model.   The following section is devoted to
phenomenology. It is shown that the inequality in sec.
(\ref{inequality}) is badly violated if the Roper resonance and
${\rm N}^*(1710)$ are  both identified as ideally mixed
pentaquarks. The various scenarios enumerated above are discussed
with a particular emphasis on the possibility of verifying or
ruling out particular scenarios via the study of pion-nucleon
scattering.

\section{The Jaffe-Wilczek Model \label{JW}}

The model is based on the following dynamical assumptions: a)  The
states can be well described in terms of a constituent quark
model with four quarks and one anti-quark.  b)  There is a strong
quark-quark interaction that binds the quarks into scalar diquarks
which are antitriplets in both color and flavor.  To be more
precise, this means that the wave function of the pentaquark is
completely dominated by configurations in which the quarks are
grouped into some particular lowest diquark configuration; this
lowest mass diquark is a spatial scalar and has $\overline{3}$
quantum numbers in both flavor and color.   c)  In the absence of
SU(3) flavor violations the lowest-lying states are an SU(3)
flavor octet and a decuplet which are nearly degenerate; all other
states are well separated. These multiplets are both $(1/2)^+$
states. d) The dynamics is such that the SU(3) symmetry breaking
Hamiltonian acting among these states is given to good
approximation by
\begin{equation}
H_I = \left ( a(n_s + n_{\overline{s}}) + b n_s  \right (2 m_s -
m_u -m_d) \; ,  \label{H}
\end{equation}
which leads to a nearly ideal mixing  in the sense that the
eigenstates of the Hamiltonian have nearly well-defined
expectation values for the number of strange quarks and strange
anti-quarks separately.  Note the form of this interaction is
slightly different from that given in ref.~\cite{JW}.  This
different form, although equivalent to that in ref. \cite{JW}, is
useful as it manifests the fact that interaction only mixes the
two multiplets via explicit SU(3) symmetry breaking.

It is worth remarking at the outset that any of these assumptions
may be questioned. The intent of the present paper is not to try
to understand the theoretical underpinnings of the model or test
whether they can ultimately be justified.   Rather, the approach
is to take the model seriously and see whether the phenomenology
that emerges is qualitatively consistent with what is observed in
nature.

The focus here is on the ideally mixed states with nucleon quantum
numbers. The first step in analyzing the decays of these states
is the construction of the two states in the context of the
model. We first accept assumptions a) and b) as stated above and
write model wave functions for a system with two scalar diquarks
(in a $\overline{3}$ flavor representation and a $\overline{3}$
color representation) and one anti-quark. The analysis begins in
the exact SU(3) limit; the form of ideal mixing as given by the
model will be computed. In general the wave function has color,
flavor, spin and spatial pieces. Explicit color will be suppressed
since all states are fully anti-symmetric with respect to color.
The flavor wave functions are expressed in terms of the
$\overline{3}$ diquarks which couple in the same way as
anti-quarks and are denoted in the obvious way:
\begin{eqnarray}
\overline{U} & \equiv & (ds) \nonumber \\
\overline{D} & \equiv & (us) \nonumber \\
\overline{S} & \equiv & (ud) \; .
\end{eqnarray}
The possible flavor wave functions with $I=1/2$ and $I_3=+1/2$ are
given by
\begin{widetext}
\begin{eqnarray}
|\overline{10}\rangle_{\rm flavor} & = & \sqrt{\frac{1}{3}} \left
( |\overline{S} \, \overline{D} \, \overline{s}\rangle +
|\overline{D} \, \overline{S} \, \overline{s}\rangle +
|\overline{S}\, \overline{S} \,
\overline{d} \rangle \right ) \nonumber \\
|8 , {\rm SA}\rangle_{\rm flavor} & = & \sqrt{\frac{1}{6}} \left
( |\overline{S} \, \overline{D} \, \overline{s}\rangle +
|\overline{D} \, \overline{S} \, \overline{s}\rangle - 2
|\overline{S} \, \overline{S} \,
\overline{d} \right ) \nonumber \\
|8 , {\rm AS}\rangle_{\rm flavor} & = & \sqrt{\frac{1}{2}} \left
( |\overline{S} \, \overline{D} \, \overline{s}\rangle -
|\overline{D} \, \overline{S} \, \overline{s}\rangle \right )
\label{flavor}
\end{eqnarray}
where we have distinguished the two possible ways of forming an
octet: symmetric-antisymmetric (SA) where the diquarks are
symmetric with each other and antisymmetric with the antiquark,
and antisymmetric-symmetric (AS) where the diquarks are
antisymmetric with each other and symmetric with the antiquark.
The most general form for the full wave functions with nucleon
quantum numbers are
\begin{eqnarray}
|\overline{10}\rangle & = & |\overline{10}\rangle_{\rm flavor}
\otimes |\overline{10}\rangle_{\rm space-spin} \nonumber
\\ | 8\rangle & = & C_{SA} \, |8 \,  SA \rangle_{\rm flavor}
\otimes |8 \, SA \rangle_{\rm space-spin} +  C_{AS} \, |8 \,  AS
\rangle_{\rm flavor} \otimes |8 \, AS \rangle_{\rm space-spin} \;
\; {\rm with} \; |C_{SA}|^2 - |C_{AS}|^2 = 1 \; . \; \label{wf}
\end{eqnarray}
\end{widetext}
The spin-space wave functions are labeled by the flavor
configuration with which they are associated.  Note that
$|\overline{10}\rangle_{\rm space-spin}$ and $|8 \, SA
\rangle_{\rm space-spin}$ must have an anti-symmetric relative
spatial wave function between the two diquarks, presumably a
p-wave.  The reason for this is simply that the diquarks are
bosons and their relative wave function must be fully symmetric.
The color part is antisymmetric implying that the spin-flavor
part is also anti-symmetric.  For the decuplet and the SA octet,
the flavor is anti-symmetric forcing the space to be
anti-symmetric.  The AS octet, however, has a symmetric spatial
wave function between the diquarks which in turn implies that the
anti-quark is in a p-wave relative to them.

It is immediately apparent that all the  matrix elements of the
number operator for strange quarks ($n_s$) are equal to the number
operator for anti-strange quarks ($n_{\overline{s}})$ when
evaluated between states with nucleon quantum numbers.  Ideal
mixing implies the states are eigenstates of the strange quark
number operators with eigenvalues of zero or unity. These states
are to be constructed from linear combinations of the states in
eq. (\ref{wf}).  Consider the linear combination with eigenvalue
zero: \begin{widetext}
\begin{eqnarray}
0 & = &n_s \left ( \alpha |\overline{10} \rangle + \beta | 8
\rangle \right ) \nonumber
\\ & = & \left ( |\overline{S} \, \overline{D} \, \overline{s}\rangle +
|\overline{D} \, \overline{S} \, \overline{s}\rangle \right )
\otimes \left ( \alpha  \sqrt{\frac{1}{3}} \, |\overline{10}
\rangle_{\rm space-spin} + \beta C_{SA}  \sqrt{\frac{1}{6}} \, |8
SA \rangle_{\rm space-spin} \right ) \nonumber \\
& + & \beta C_{AS} \left ( |\overline{S} \, \overline{D} \,
\overline{s}\rangle - |\overline{D} \, \overline{S} \,
\overline{s}\rangle \otimes |8 AS \rangle_{\rm space-spin} \right
) \label{zeroe}
\end{eqnarray}
\end{widetext}
It is easy to see that this equality cannot be satisfied unless
\begin{eqnarray}
\beta & = &-\sqrt{2} \alpha \\
C_{AS}  = 0 \; & \; & \; C_{SA}=1 \label{Ccond}\\
 |8 \, SA
\rangle_{\rm space-spin} & = & |\overline{10} \rangle_{\rm
space-spin} \; .\label{scond}
\end{eqnarray}
The conditions in eqs. (\ref{Ccond}) and (\ref{scond}) imply that
ideal mixing requires the space-spin wave functions in the two
states be identical.  This is a very strong condition on the
underlying dynamics.  These conditions will  be of some
significance in the phenomenological discussions in
sect.~(\ref{phen}).  Note that any substantial deviation from
this condition such as a large mixing of the octet pentaquark
component with ordinary three quark states will spoil ideal
mixing.

 It is easy to see that only way to construct such
states from a superposition of the states in eq.~(\ref{wf}) is:
\begin{eqnarray}
|n_s =0 \rangle   & =  & \sqrt{\frac{1}{3}} \, |\overline{10}
\rangle +
\sqrt{\frac{2}{3}}  \, |8 \rangle \nonumber \\
|n_s =1  \rangle   & = &
 \sqrt{\frac{2}{3}} \,  |\overline{10}
\rangle - \sqrt{\frac{1}{3}} \,  |8 \rangle \\
\label{mix}\end{eqnarray} subject to the constraints in eqs.
(\ref{Ccond}) and (\ref{scond}).  The state with $n_s=0$ is
identified as the lower of the two ${\rm N}^*$ states while the
$n=1$ state is higher. The construction of the explicit states
allows for the derivation of an inequality for the decay widths.
This is derived in the following section.

\section{An inequality for decay widths\label{inequality}}

Approximate SU(3) flavor symmetry has played an essential role in
our understanding of hadronic physics.  Typically the symmetry
works quite well and violations can be expected to be less than
$\sim 30\% $ for generic quantities.  However, there are cases
where the violations of SU(3) are characteristically much larger.
These include: i) masses of pseudo-Goldstone bosons; ii) SU(3)
multiplets with nearly degenerate masses as occurs in cases of
ideal mixing; iii) decay widths near threshold (even when
couplings respect SU(3) symmetry to good approximation, phase
space factors near threshold vary rapidly with small changes in
mass and can yield quite different widths).  Fortunately, these
cases are well understood.  Apart from these well-established
effects SU(3) works quite well phenomenologically.

The key to establishing an inequality for pentaquark decay widths
in the Jaffe-Wilczek model are the assumptions  that {\it all}
SU(3) violating effects are the three enumerated above, and that
the physical ${\rm N}^*$ pentaquark states are exact ideal
mixtures of the octet and the anti-decuplet. Clearly, neither of
these assumptions is strictly correct.  Thus the inequality
derived is approximate. Small violation of the inequality can
arise either from small deviations from ideal mixing or from
garden variety SU(3) violations.  Large violations, however,
indicate either large deviations from ideal mixing in
contradiction to the Jaffe-Wilczek model or a new and previously
unknown mechanism for large effects due to SU(3) breaking.  Since
the latter is quite implausible, large violations of the
inequality would strongly suggest that the model is untenable.

The only additional assumption about the dynamics which goes into
the derivation is that the decay of an unstable particle can be
cleanly treated via the coupling of the unstable particle to the
continuum of open channels with no significant interference from
background effects.  In this circumstance, the partial width for
the two-body decay of a $(1/2)^+$ baryon ($B'$) into another
baryon ($B$) plus a pseudo-scalar meson ($P$) may be computed by
combining a coupling constant parameterizing the strength with
appropriate kinematic factors which incorporates both the phase
space and the p-wave nature of the coupling\cite{SZ} of a
pseudo-scalar meson to $(1/2)^+$ states.
\begin{eqnarray}
&{}&\Gamma_{B' \rightarrow B P}  =  |g_{P B' B}|^2 \, \kappa_{P B' B}
\nonumber \\
&{}& \kappa_{P B' B} =  \frac{q}{4 \pi M_{B'}} \left ( \sqrt{q^2
+ M_B^2} -M_B \right ) \label{width}
\end{eqnarray}
where the kinematical factor $\kappa_{P B' B}$ depends on the
momentum of the decaying fragments in the rest frame of $B'$:
\begin{equation}
M_{B'} = \sqrt{q^2 + M_{P}^2} +  \sqrt{q^2 + M_{B}^2}
\end{equation}.

Here we relate the decays of three baryons, the $\theta^+$ (into a
kaon plus a nucleon) and the two ideally mixed pentaquarks with
nucleon quantum numbers (into a pion plus a nucleon). The ideally
mixed  states will be labeled $N_0$ and $N_1$ where the subscript
indicates the number of strange quarks (which equals the number
of anti-strange quarks in the state) so that with the tentative
identification of Jaffe and Wilczek, the $N_0$ is the Roper
resonance and the $N_1$ is the ${\rm N}^*(1710)$.  Given the
assumption that the only sources of SU(3) violation are due to
phase space, ideal mixing, and the masses of pseudo-scalar mesons,
it follows from eq.~(\ref{mix}) that
\begin{eqnarray}
g_{\pi N_1  N} & = & - \sqrt{\frac{1}{3}} g_{\pi N_8 N} +
\sqrt{\frac{2}{3}} g_{\pi N_{\overline{10}} N }\nonumber \\
 g_{\pi N_0  N} & = &  \sqrt{\frac{2}{3}} g_{\pi N_8 N} +
\sqrt{\frac{1}{3}} g_{\pi N_{\overline{10}} N }\label{gr}
\end{eqnarray}
where $g_{\pi N_8 N}$ and $g_{\pi N_{\overline{10}} N}$ are the
coupling constants for the unmixed pure SU(3) states with nucleon
quantum numbers. Simple algebra yields the inequality:
\begin{widetext}
\begin{equation}
\left | |g_{\pi N_0  N}|^2 - 2 |g_{\pi N_1  N}|^2 + |g_{\pi
N_{\overline{10}} N }|^2  \right | \le 2 \sqrt{2} |g_{\pi
N_{\overline{10}} N }| \sqrt{|g_{\pi N_0  N}|^2 +  |g_{\pi N_1
N}|^2 -|g_{\pi N_{\overline{10}} N }|^2 } \, ,\label{ne1}
\end{equation}
where (\ref{ne1}) is an inequality rather than equality due to
unknown relative phases between the coupling constants.  The
important thing is that this inequality is of a form which
relates the $N_0$ and $N_1$ couplings to the $N_{\overline{10}}$
coupling, which by SU(3) symmetry, is related to the coupling
constant for the decay of the $\theta^+$ into kaon plus nucleon:
\begin{equation}
g_{\pi N_{\overline{10}} N } = \frac{1}{2}g_{K \theta^+ N }
\label{grel}\end{equation} where the factor of 1/2 is an SU(3)
isoscalar factor\cite{Ds}.  Combining eqns.~(\ref{width}),
(\ref{ne1}) and (\ref{gr}) yields
\begin{equation}
\left | \frac{\Gamma_{\pi N_0  N}}{\kappa_{\pi N_0  N}} - 2
\frac{\Gamma_{\pi N_1  N}}{\kappa_{\pi N_1  N}} + \frac{\Gamma_{K
\Theta^+ N }}{4 \kappa_{K \Theta^+ N }} \right | \le
 \sqrt{ \left( \frac{2
\Gamma_{K \Theta^+ N }}{ \kappa_{K \Theta^+ N }}\right ) \left(
\frac{\Gamma_{\pi N_0 N}}{\kappa_{\pi N_0 N}} + \frac{\Gamma_{\pi
N_1 N}}{\kappa_{\pi N_1 N}} - \frac{\Gamma_{K \Theta^+ N }}{4
\kappa_{K \Theta^+ N }} \right )} \, , \label{ne2}
\end{equation}
\end{widetext}
where the $\kappa$ are the kinematic factors defined in eq.
(\ref{width}).

Inequality (\ref{ne2}) is the principal result of this work.  It
constrains the possible widths of the various decays and depends
only upon the assumptions outlined above.

\section{Phenomenological implications\label{phen}}

Inequality (\ref{ne2}) can be used to rule out the possibility
that $N_0$ is the ${\rm N}^*(1440)$ ({\it i.e.}, the Roper
resonance) and $N_1$ is  the ${\rm N}^*(1710)$.  It can be seen
by taking the  masses and partial widths of these states from the
particle data table\cite{pdg}.  As always, there is some ambiguity
in extracting these from experiment.  The quoted values for the
mass of the ${\rm N}^*(1440)$ is in a range from 1430-1470 MeV
with a best estimate of $\approx 1440$ MeV; the total width is
estimated to be in the range from 250-450 MeV with a best
estimate of $\approx 350$ MeV; and the branching fraction to $\pi
\, N$ is 60\%-70\%.  The quoted values for the mass of the ${\rm
N}^*(1440)$ is in a range from 1680-1740 MeV with a best estimate
of $\approx 1710$ MeV; the total width is estimated to be in the
range from 50-250 MeV with a best estimate of $\approx 100$ MeV;
and the branching fraction to $\pi \, N$ is 10\%-20\%.

As a first test, associate the Roper resonance with $N_0$, the
${\rm N}^*(1710)$ with $N_1$ and take the best estimate values
for the masses and partial widths: $M_{N_1} =1440$ MeV,
$\Gamma_{\pi N_1 N}=227.5$ MeV;  $M_{N_0} =1710$ MeV,
$\Gamma_{\pi N_0 N}=15$ MeV.  Taking these values and using the
direct experimental bound of 20 MeV for $\Gamma_{K \Theta^+ N }$,
one sees that the left-hand side of the inequality (\ref{ne2}) is
71.8 while the right-hand side is 61.8.  This violates the
inequality, but only modestly; as noted above, modest violations
can be understood as arising for ordinary SU(3) mixing  or small
deviations from ideal mixing.  However, if the indirect bound on
the width of the $\theta^+$ of 2 MeV from
ref.~\cite{width2,width3} is used as the $\theta^+$ width, the
left-hand side of the inequality is 77.1 MeV while the right-hand
side is 19.7.  The inequality is violated by nearly a factor of
4.  This is a gross violation and clearly indicates that
something is wrong. Of course, the 2 MeV is  a {\it bound}; the
actual width could be smaller.  If, for example, the width were 1
MeV, the violation would be even worse with the left-hand side a
factor of nearly 6.

The violations of the inequality reported in the previous
paragraph were based on  best estimate values for the masses and
widths for the Roper resonance and the ${\rm N}^*(1710)$, so these
values are uncertain.  Is it possible that the inequality would
be satisfied or nearly satisfied if the masses and widths for
these states had more fortuitous values?  To test this we can
take the mass of the Roper resonance at the top of its plausible
range (1470 MeV), the width at the bottom of its range (250 MeV),
and the branching fraction to the pion-nucleon channel at the
bottom of its range (60\%) while taking the mass of the ${\rm
N}^*(1710)$ at the bottom of its range (1680 MeV), its width at
the top of its range (250 MeV), and the branching fraction at the
top of its range (10\%).  Making these extremely optimistic
assumptions one finds  the inequality remains badly if one takes
the width of the $\theta^+$ to be 2 MeV, as suggested by the
indirect bound in ref.~\cite{width2,width3}: the left-hand side is
33.5 while the right-hand side is 15.8 for a violation of more
than  a factor of 2. If the width of the $\theta^+$ is taken to
be 4 MeV---which is quite conservative---the violation is still
by a factor of 1.5.

The reason for the violations is quite clear.  The very small
width of the $\theta^+$ implies that the coupling of the
$\overline{10}$ state is extremely small and thus the decay of
both the $N_1$ and $N_0$ states are either both quite narrow or
essentially dominated by the octet contribution. However, the
Roper resonance width is large, so by making the identification of
the Roper resonance with the $N_0$, we fix the octet coupling to
be large. This in turn implies that the $N_1$ state must also be
dominated by the octet and have large partial width to pion plus
nucleon. However, the ${\rm N}^*(1710)$ has a very narrow partial
width. The inequality simply codifies this.

Violation of the inequality depends critically on the bound
deduced for the width of the $\theta^+$.  Thus it is of great
significance to understand the reliability of this bound. The
basic point is quite simple: $\theta^+$ decays into $K^+ N$ and
hence corresponds to a resonance in $K^+ N$ scattering. Of course,
we have no neutron targets, but the deuteron is loosely bound and
may be treated effectively as a neutron plus a proton. The wave
function of the deuteron effectively smears the neutron's inital
momentum. Moreover, the height of the $K^+ N$ resonance is fixed
by unitarity since the kinematics are such that no inelastic
channels are open.  Thus the area under the $K^+ D $ curve is
simply proportional to the width of the $\theta^+$. Since the wave
function smearing of the deuteron is known to be significantly
larger than the the width of  the $\theta^+$, the $K^+ D$
scattering effectively counts the total strength under the $K^+
N$ resonant peak which is proportional to the
width\cite{width1}.  However, in $K^+ D$ there is no observed
structure in the vicinity of the $\theta^+$ allowing one to bound
the width\cite{width1}.  A rather crude limit of 6 MeV for the
width was reported  based on a study of a rather limited set of
data. More complete data  reduces the bound to 1.5 MeV
\cite{width3}. This is consistent with the bound of 1-2 MeV found
from a full re-analysis of the $K^+$ scattering data
base\cite{width2}.  A similar analysis of the xenon experiment
\cite{width4} produces a bound of 1 MeV, although the nuclear
structure uncertainties with Xe are clearly larger than with the
deuteron. Together, these analyses lead to a fairly conservative
estimate of 2 MeV as an upper bound for the width.

It is clear that if the $\theta^+$ width is as narrow as reported
in  refs.~\cite{width2,width3}, then the Jaffe-Wilczek model with
the mixed states identified as the Roper resonance and the ${\rm
N}^*(1710)$ is simply not viable.  One possibility, of course, is
that the Jaffe-Wilczek model is wrong. The other is that the
identification of the two ideally mixed states with the Roper
resonance and the ${\rm N}^*(1710)$ is not correct. Let us
explore this second possibility in some detail. There are three
scenarios to consider: i) the ${\rm N}^*(1710)$ state is, in fact,
a nearly ideally mixed pentaquark and it has a previously
undiscovered narrow partner with a mass of  approximately $\rm
1450$ MeV; ii) the Roper resonance is a nearly ideally mixed
pentaquark and it has a previously undiscovered broad partner
with a mass of approximately $\rm 1700$ MeV;  iii) neither the
lower nor the upper ${\rm N}^*$ states have been detected.

First consider scenario i).  If one takes the  ${\rm N}^*(1710)$
as the $N_1$ and assumes that the there exists a previously
undetected ${\rm N}^*$ state around 1450 MeV, then inequality
(\ref{ne2}) can be used to get at least crudely bound its width.
Taking the best estimate values for the ${\rm N}^*(1710)$, a
$\theta^+$ width of 2 MeV, and assuming the inequality is not
violated  gives an approximate upper bound for the partial width
of this undetected state into pion plus nucleon of 35 MeV. In
fact, if this state does exist it probably must be much narrower
than this to have not previously been observed. An analysis
similar to the one recently done in ref.~\cite{gw} can place
possible bounds on the width of any heretofore unknown state from
pion-nucleon scattering. An optimistic view is that this analysis
might identify a candidate for such a state.  The possibility
that a light state with nucleon quantum numbers which is
sufficiently narrow to have escaped detection to date has been
considered by Gothe and Nussinov\cite{GN}. Clearly the prediction
of an extremely narrow new resonant state in pion-nucleon
scattering is a remarkable prediction and, if seen, would lead to
great credence of the Jaffe-Wilczek model. There is a possible
theoretical objection to this scenario: it requires two
resonances with the same quantum numbers and virtually  the same
mass; namely, the Roper resonance and the new state. This is
unprecedented in hadronic physics and might be viewed as quite
unnatural.

Next consider scenario ii).  If one takes  the ${\rm N}^*(1440)$
as the $N_0$ and assumes that  there exists a previously
undetected ${\rm N}^*$ state around 1700 MeV, then inequality
(\ref{ne2}) can be used to at least crudely bound its width.
Taking the best estimate values for the ${\rm N}^*(1440)$, a
$\theta^+$ width of 2 MeV, and assuming the inequality is not
violated  gives a lower bound for the partial width of this
undetected state into pion plus nucleon of 245 MeV.  The
interesting phenomenological question is whether there can exist
a state in this region which has not been previous observed.
Again it would be of use to do a detailed study of the data base
of pion-nucleon scattering to see what bounds can be placed on
such a state.  The concern is simply that such a state might be
so wide and so inelastic that there is no way to distinguish such
a resonance from a smooth background.  Again there are
theoretical concerns with this scenario.  Note the same issue
arises as in scenario i) that there would be two states with the
same quantum numbers which strongly overlap and, as noted before,
is unprecedented in hadronic physics.  This issue may not be as
serious in this case: the ${\rm N}^*(1710)$ is only a three-star
resonance in the particle data book\cite{pdg} and its existence
may be questioned.

There is another theoretical issue which makes scenario ii) rather
problematic, namely, the existence of two widely different width
scales in the problem, a very wide octet pentaquark, and a very
narrow anti-decuplet. Note that this same issue was confronted in
ref. \cite{JW} where an attempt was made to reconcile the narrow
$\theta^+$ with the wide Roper resonance. Moreover, with the
recent determination that the $\theta^+$ width is far smaller
than the experimental bound, the problem has become even more
acute. Jaffe and Wilczek argued in ref.~\cite{JW} that the
internal and flavor structure in the Roper resonance were quite
different from that of the $\theta^+$ and that this difference
could account for the differing widths.  In fact, this is not
completely correct. As shown in sec.~\ref{JW}, ideal mixing can
occur only if the pentaquark wave function for the octet and the
anti-decuplet are identical except for their flavor structure;
they must have the same internal wave functions in terms of spin
and space.    At least naively one would think it unlikely that
the octet would have a very large coupling to the continuum while
an octet state of essentially the same internal structure has a
very small coupling.

It is not clear how seriously one ought to regard this problem.
The key issue is whether the problem is serious enough to rule out
scenario ii). Glozman has taken the position  that the disparate
widths for the $\theta^+$ and the Roper resonance rule out the
possibility that the Roper resonance is a pentaquark \cite{Gl}.
This position is probably too strong.  It certainly does seem
perverse to suggest that  both the widest and the narrowest
baryon resonances known in this region of the baryon spectrum are
states with virtually identical spin-space wave functions.
However, there appears to be no mathematically rigorous argument
which rules out this possibility. Nevertheless the widely
differing widths make this scenario at the very least a somewhat
unattractive possibility.

The third scenario is that neither of the two states has been
seen heretofore.  If this is true then both states are very
narrow.  The inequality requires either both states be wide or
both be narrow so the possibility of two broad states near 1450
MeV (the Roper resonance and a new state) can be ruled out.   In
some ways this scenario is unattractive since it depends on the
existence of two states at low energies which have evaded
detection until now. However, this possibility cannot be totally
disregarded because the ${\rm N}^*(1710)$ is not completely well
established (it is listed as a three-star resonance in the
particle data book, and not seen at all in a recent re-analysis
of the pion-nucleon scattering data\cite{gw2} while possible
candidates for narrow resonances were seen at 1680 MeV and 1730
MeV \cite{gw}).

It is  worth noting that these conclusions hold rather more
generally than for the Jaffe-Wilczek model.  Any model which
predicts nearly ideal mixing and positive parity pentaquarks will
face the same constraints as the Jaffe-Wilczek model. All such
models will similarly face the problem that the physical states in
the theory with nucleon quantum number cannot be identified with
the ${\rm N}^*(1440)$ and  ${\rm N}^*(1710)$. Thus, any such model
will require the existence of at least one presently unknown
nucleon resonance in a region which has been well explored in the
past.

To summarize: the extreme narrowness of the $\theta^+$ as
extracted from the absence of a detectable resonance
kaon-deuteron scattering, strongly constrains the phenomenology
of the Jaffe-Wilczek model.  Given this narrowness, the
identification of the nearly ideally mixed nucleon states as the
Roper resonance and the ${\rm N}^*(1710)$ is not tenable.  To be
viable the model requires that either both ideally mixed states
be narrow or both wide, and there is at least a theoretical
prejudice that the scenario where they are both narrow is
preferred.  A re-analysis of pion-nucleon scattering in the
region around 1450 MeV for evidence of a previously missed narrow
resonance may shed light on the situation.

\section*{Acknowledgments}  The author thanks S.~H.~Lee and
S.~Nussinov for many interesting discussions.  The author also
thanks R.~L.~ Jaffe for a useful communication. The support of the
U.S. Department of Energy under grant DE-FG02-93ER-40762 is
gratefully acknowledged.

\end{document}